\documentclass[
    reprint,          % use for final journal-like layout
    aps,              % APS journal style
    prb,              % Physical Review B
    amsmath,amssymb,  % math packages
    superscriptaddress% authors' affiliations superscript
]{revtex4-2}

\usepackage[T1]{fontenc}
\usepackage[utf8]{inputenc}
\usepackage{graphicx}   % Include figure files
\usepackage{xcolor}     % Color for text or figures
\usepackage{hyperref}   % Hyperlinks for references, figures, etc.
\usepackage{float}
\usepackage{placeins}

\hypersetup{
    colorlinks=true,
    linkcolor=blue,
    citecolor=blue,
    urlcolor=blue
}

\begin{document}

\title{Structural dependence of quantum transport properties \\
       on topological nodal-line semimetal bilayer borophene}

\author{C. J. P\'aez-Gonz\'alez}
\email{cjpaezg@uis.edu.co}
\affiliation{School of Physics, Universidad Industrial de Santander, 680002, Bucaramanga, Colombia.}

\author{C. E. Ardila-Guti\'errez}
\affiliation{School of Physics, Universidad Industrial de Santander, 680002, Bucaramanga, Colombia.}

\author{D. A. Bahamon}
\email{dario.bahamon@mackenzie.br}
\affiliation{School of Engineering, Mackenzie Presbyterian University, S\~ao Paulo - 01302-907, Brazil}
\affiliation{MackGraphe – Graphene and Nanomaterials Research Institute, Mackenzie Presbyterian University, S\~ao Paulo -01302-907, Brazil}

\date{\today}

\begin{abstract}
This work presents the electronic and transport properties of bilayer borophene nanoribbons. In the first part, a four-orbital tight-binding model is derived by fitting the \emph{ab initio} band structure. The transport properties of armchair and zigzag bilayer borophene nanoribbons are then analyzed, both with and without periodic boundary conditions. In both scenarios, the nodal line causes conductance to increase with width and exhibit oscillations in narrow nanoribbons. Additionally, plots of current and charge density reveal that edge states have a more pronounced impact in narrower nanoribbons. Finally, uniaxial tensile strain is introduced as a tool to engineer the number of available transport channels.
\end{abstract}

\maketitle

\section{Introduction\label{sec1}}

The exfoliation of single-layer graphene from multilayer graphite systems in 2004~\cite{novoselov_electric_2004} marked the beginning of a surge in interest in two-dimensional materials. Since then, a growing number of two-dimensional materials have been discovered, including silicene~\cite{liu_various_2014}, germanene~\cite{liu_multiple_2015}, borophene~\cite{mannix_synthesis_2015}, and transition metal dichalcogenides (TMDs)~\cite{manzeli_2d_2017}. These materials display a wide range of novel properties—mechanical, thermal, electrical, magnetic, and topological—while also presenting several experimental challenges. As silicon is one of carbon's closest neighbors, it typically undergoes \( sp^3 \) orbital hybridization, leading to the instability of the silicene structure~\cite{jackson_unraveling_2004}. Germanene also presents preparation challenges due to the larger atomic spacing between its nearest neighbors~\cite{davila_germanene_2014}. Similarly, bilayer borophene (\(\mathcal{BB}\)) cannot be produced via mechanical exfoliation like graphene, as bulk boron does not naturally occur. Instead, it is typically grown using molecular beam epitaxy under controlled conditions on various metal substrates, such as silver~\cite{liu_borophene_2022}, gold~\cite{kiraly_borophene_2019}, copper~\cite{chen_synthesis_2022, wu_large-area_2019}, and ruthenium~\cite{sutter_large-scale_2021}. This method has yielded materials with either metallic~\cite{zhong_electronic_2018} or semiconducting~\cite{zhou_semimetallic_2014, mohebpour_transition_2022, yan_semiconducting_2023} properties, along with great mechanical stability~\cite{ma_graphene-like_2016, pang_super-stretchable_2016, sun_two-dimensional_2017} and enhanced resistance to oxidation~\cite{gao_structure_2018}. This stability is attributed to the boron-boron bonds connecting both layers~\cite{qureshi_bilayer_2024}.

Several phases of \(\mathcal{BB}\), including AA, AB, and AB' stackings, have been primarily investigated through \emph{ab initio} methods~\cite{zhong_electronic_2018, gao_structure_2018, scopel_bridging_2023, yan_bottom-up_2022}. Among these, AA stacking, which is the focus of this study, has garnered significant attention due to its stability and unique properties~\cite{li_topological_2020, wang_second-order_2024, yang_optical_2022, gao_density_2023, cheng_giant_2022, soskic_first-principles_2024, yan_prediction_2021, penev_can_2016, zhao_superconductivity_2016}. These properties stem from the presence of two non-hybridizing Dirac cones~\cite{crasto_de_lima_orbital_2019, gupta_dirac_2018}, which form a non-flat nodal line~\cite{nakhaee_dirac_2018, xu_ideal_2019} with non-trivial Berry curvature~\cite{xiao_berry_2010} and a Chern number of \(C = 2\)~\cite{nakhaee_dirac_2018, xu_ideal_2019}.  

Building on this discussion, this study explores the electronic and transport properties of \(\mathcal{BB}\) nanoribbons with both armchair and zigzag edge orientations. To achieve this, the hopping and energy parameters of the tight-binding Hamiltonian for \(\mathcal{BB}\) were first determined using the Slater-Koster (SK) method in the TBStudio program~\cite{nakhaee_tight-binding_2020}. Subsequently, the tight-binding model was implemented in the Kwant library~\cite{groth_kwant_2014} to investigate quantum transport in \(\mathcal{BB}\) nanoribbons (armchair and zigzag) with varying widths. Our results demonstrate that the conductance of narrow nanoribbons oscillates at the charge neutrality point as a function of width. To better understand this effect, we studied cases with and without periodic boundary conditions along the width direction. These studies revealed that the oscillations are a direct consequence of the nodal line and the quantization of transverse momentum, a phenomenon not observable in other bilayers where only a Dirac point exists~\cite{BahamonChV3,de_castro_fast_2023,paez2014,de_castro_fast_2023b}. With hard wall boundary conditions, edge states also emerge. In these systems, conductance oscillations are still present; however, due to coupling effects, it is not possible to derive a mathematical expression for the conductance oscillations.

This paper is organized as follows: Section~\ref{sec2} describes the computational details, including the density functional theory (DFT) calculations and the development of the tight-binding model. Section~\ref{sec3} presents the results and discussion, addressing the transport properties, the effects of Brillouin zone folding, and the impact of edge-induced transport alterations. Section~\ref{sec_strain} explores the impact of a 5\% uniaxial tensile strain, and Section~\ref{sec4} concludes the study. 

%%%%%%%%%%%%%%%%%%%%%%%%%%%%%%%%%%%%%%%%%%%%
\section{Computational Details: Density Functional Theory and Tight-Binding\label{sec2}}
%%%%%%%%%%%%%%%%%%%%%%%%%%%%%%%%%%%%%%%%%%%%

%%%%%%%%%%%%%%%%%%%%%%%%%%%%%%%%%%%%%%%%%%%%
\begin{figure}[t]
    \centering
        \includegraphics[width=1\columnwidth]{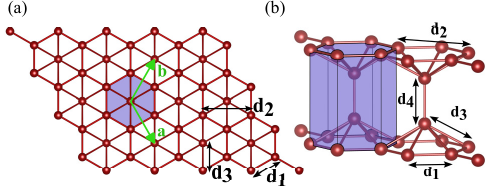}
    \caption{\(\mathcal{BB}\) structure: (a) top view and (b) side view. The shaded area indicates the primitive cell. Hopping distances $d_{i}$ are shown by black arrows. The lattice vectors \(\mathbf{a}\) and \(\mathbf{b}\) are in green. Comparison of DFT (blue) and tight-binding (yellow) band structures is presented in (c).}
    \label{fig1}
\end{figure}
%%%%%%%%%%%%%%%%%%%%%%%%%%%%%%%%%%%%%%%%%%%%

The crystalline structure of \(\mathcal{BB}\) is shown in Fig.~\ref{fig1} from two perspectives. In Fig.~\ref{fig1}(a), a top view illustrates the hexagonal unit cell containing six boron atoms. After optimizing the atomic positions, the interatomic distances were found to be \(d_1 = 1.65\,\text{\AA}\), \(d_2 = 2.85\,\text{\AA}\), \(d_3 = 1.89\,\text{\AA}\), and \(d_4 = 1.70\,\text{\AA}\). The lattice constants were determined as \(|\mathbf{a}| = |\mathbf{b}| = 2.85\,\text{\AA}\), with an angle of \(\theta_{\mathrm{ab}} = 120^\circ\)~\cite{ma_graphene-like_2016}. Fig.~\ref{fig1}(b) presents a side view of the bilayer, revealing two interlocking honeycomb lattices connected by vertical bonds.

To establish an accurate model for extracting the tight-binding parameters for \(\mathcal{BB}\), DFT calculations were performed using \textsc{Quantum ESPRESSO} \cite{scandolo_first-principles_2005}. A supercell method with a plane-wave basis set~\cite{kresse_efficiency_1996} and the plane-wave pseudopotential method were employed for structural relaxations and electronic band structure calculations. The exchange-correlation energy was computed using the generalized gradient approximation formulated by Perdew, Burke, and Ernzerhof~\cite{perdew_generalized_1996}. A cutoff energy of 400~eV was chosen, balancing computational efficiency with the required precision for accurate electronic state modeling~\cite{nakhaee_dirac_2018}. Atomic positions were optimized until a force threshold of less than 25~meV/\AA{} was achieved, ensuring structural stability for reliable predictions of electronic properties. For Brillouin zone sampling, a (12~$\times$~12~$\times$~1) $k$-point mesh was employed, chosen for its effectiveness in detailed electronic band structure analysis; for examinations along high-symmetry directions, a refined $k$-point mesh was utilized.

The resulting DFT band structure is shown by the violet lines in Fig.~\ref{fig2}. The blue box highlights two unhybridized Dirac cones, located at the \(K\) point. The nodal line is formed by the intersection of these cones in the 2D momentum space~\cite{xu_ideal_2019}. Since the plot follows the \(\Gamma\)--\(K\)--\(M\)--\(\Gamma\) path, the nodal line is intersected twice at points \(\alpha\) and \(\alpha'\). The energies at these points differ due to the trigonal warping effect, which induces angular variations in the energy of the nodal line~\cite{nakhaee_dirac_2018}.

% =================================================
\subsection{The Optimized Tight-Binding Model Hamiltonian\label{sec2.1}}
% =================================================

%%%%%%%%%%%%%%%%%%%%%%%%%%%%%%%%%%%%%%%%%%%%
\begin{figure}[t]
    \centering
        \includegraphics[width=1\columnwidth]{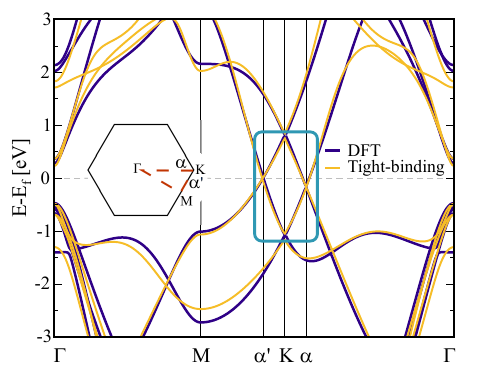}
    \caption{ Comparison of DFT (blue) and tight-binding (yellow) band structures.}
    \label{fig2}
\end{figure}
%%%%%%%%%%%%%%%%%%%%%%%%%%%%%%%%%%%%%%%%%%%%

The tight-binding Hamiltonian can be written as:

\begin{equation}
\label{eq:H}
H=\sum_{j\mu,\,i\nu } t_{j \mu, i \nu}\, c_{i\nu}^{\dagger} c_{j\mu}+\sum_{i\nu} \varepsilon_{i\nu}\, c_{i\nu}^{\dagger} c_{i\nu},
\end{equation}

\noindent where the operator $c_{i\nu}^{\dagger}$($c_{j\mu}$) creates (annihilates) one electron on the orbitals $\nu(\mu) =s,~p_x,~p_y,~\text{and}~p_z$ at sites \(i\)(\(j\)). The hopping parameters \( t_{j \mu, i \nu} \) are written as a function of the Slater-Koster (SK) parameters:

\begin{equation}
\begin{aligned}
t_{j s, i s}&=\langle s^j|H| s^i\rangle =V_{ss\sigma}, \\
t_{j s, i p_\gamma}&=\langle s^j|H| p^i_\gamma \rangle =n_\gamma V_{sp\sigma}, \\
t_{j p_\zeta, i p_\gamma}&=\langle p^j_\zeta|H| p^i_\gamma\rangle =(\delta_{\zeta\gamma}-n_\zeta n_\gamma) V_{pp\pi}+n_\zeta n_\gamma V_{pp\sigma},
\end{aligned}
\end{equation}
where \( n_\gamma = \mathbf{r} \cdot \mathbf{e}_\gamma / |\mathbf{r}| \) represents the directional cosine, with \( \mathbf{r} \) as the bond vector and \( \gamma \) encompassing \( x \), \( y \), and \( z \). Here, \( \varepsilon_{i\nu} \) denotes the on-site energy for the orbital \( \nu \) of the \( i \)-th atom, and \( V_{ss\sigma} \), \( V_{sp\sigma} \), \( V_{pp\sigma} \), and \( V_{pp\pi} \) are the Slater-Koster parameters. The SK parameters were fitted using the TBStudio program~\cite{nakhaee_tight-binding_2020}; during this process, the Fourier transform of the tight-binding Hamiltonian was iteratively adjusted to match the DFT band structure with high precision. Specifically, the process stopped when the mean squared error reached a value of 0.001. The resulting band structure, shown in yellow in Fig.~\ref{fig2}, demonstrates that the tight-binding Hamiltonian accurately reproduces the DFT band structure for Fermi energies between \(-1\) and \(1\) eV. The parameters used for this plot are listed in Table \ref{table:TBpara}. The same table also includes the hopping parameters under 5\% uniaxial tensile strain, which will be utilized in Section \ref{sec_strain}.

%%%%%%%%%%%%%%%%%%%%%%%%%%%%%%%%%%%%%%%%%%%%
% Table: Hoppings and On-site Energies
\begin{table*}[!ht]
\centering
\begin{tabular}{lcccccccccccc}
      & \multicolumn{4}{c}{Unstrained} & \multicolumn{4}{c}{X-strain=5\%} & \multicolumn{4}{c}{Y-strain=5\%} \\
      & $V_{ss\sigma}$ & $V_{sp\sigma}$ & $V_{pp\sigma}$ & $V_{pp\pi}$ 
      & $V_{ss\sigma}$ & $V_{sp\sigma}$ & $V_{pp\sigma}$ & $V_{pp\pi}$ 
      & $V_{ss\sigma}$ & $V_{sp\sigma}$ & $V_{pp\sigma}$ & $V_{pp\pi}$ \\
\hline
$d_{1}$ & -34.198 & 7.863 & 3.472 & -2.447 
       & -17.128 & 11.685 & 6.444 & -3.595 
       & -53.540 & 19.853 & 9.036 & -2.706 \\
$d_{2}$ & 16.855 & -2.405 & -0.412 & -0.088 
       & 5.928 & 1.837 & -0.316 & -0.352 
       & 22.271 & 3.605 & 1.490 & -0.608 \\
$d_{3}$ & 1.342 & 1.944 & 2.375 & -0.993 
       & -2.572 & 3.043 & 4.990 & -2.172 
       & -5.352 & 4.102 & 5.514 & -1.556 \\
$d_{4}$ & 8.677 & -0.141 & 1.484 & 1.455 
       & 7.108 & -2.360 & -2.116 & 3.299 
       & 11.643 & -1.109 & -1.944 & 3.353 \\
\hline
      & $\varepsilon_S$ & $\varepsilon_{P_y}$ & $\varepsilon_{P_z}$ & $\varepsilon_{P_x}$ 
      & $\varepsilon_S$ & $\varepsilon_{P_y}$ & $\varepsilon_{P_z}$ & $\varepsilon_{P_x}$ 
      & $\varepsilon_S$ & $\varepsilon_{P_y}$ & $\varepsilon_{P_z}$ & $\varepsilon_{P_x}$ \\
\hline
      & -4.127 & 2.535 & 4.550 & 2.361 
      & -4.620 & 4.858 & 5.801 & 4.889 
      & -9.491 & 5.224 & 5.560 & 5.467 \\
\end{tabular}
\caption{Slater-Koster parameters for all four interatomic distances, as well as the on-site energies for each orbital, are provided for systems both without strain and with 5\% uniaxial tensile strain applied in the \(X\) and \(Y\) directions. After optimization, the interatomic distances were determined to be \(d_1 = 1.69\,\text{\AA}\) and \(1.68\,\text{\AA}\), \(d_2 = 2.88\,\text{\AA}\) and \(2.90\,\text{\AA}\), \(d_3 = 1.92\,\text{\AA}\) and \(1.90\,\text{\AA}\), and \(d_4 = 1.70\,\text{\AA}\) for the \(X\)-strained and \(Y\)-strained systems, respectively. The lattice constants were found to be \(|\mathbf{a}| = |\mathbf{b}| = 2.888\,\text{\AA}\) and \(2.899\,\text{\AA}\), with angles \(\theta_{\mathrm{ab}} = 117.8^\circ\) and \(121.3^\circ\), respectively. Units are in eV.}
\label{table:TBpara}
\end{table*}
%%%%%%%%%%%%%%%%%%%%%%%%%%%%%%%%%%%%%%%%%%%%
%%%%%%%%%%%%%%%%%%%%%%%%%%%%%%%%%%%%%%%%%%%%
\begin{figure}[t]
    \centering
        \includegraphics[width=1\columnwidth]{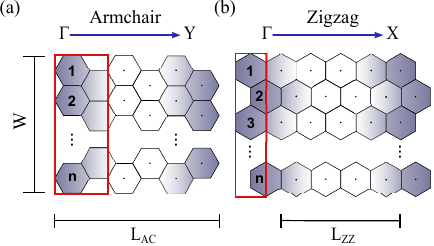}
    \caption{(a) $\mathcal{BB}$ transport configuration. A scattering region of length $L_{i}$ and width $W$ is attached to semi-infinite leads. Zigzag (a) and armchair (b) edges are shown. $W$ is measured by the number of primitive cells $n$.}
    \label{fig3}
\end{figure}
%%%%%%%%%%%%%%%%%%%%%%%%%%%%%%%%%%%%%%%%%%%%
Using the obtained tight-binding parameters, the electronic and transport properties of $\mathcal{BB}$ zigzag and armchair nanoribbons (NRs) can be analyzed. Figures~\ref{fig3}(a) and (b) depict the structure of the NRs, with the red box highlighting the unit cell. The width of the NRs is determined by the number of hexagonal plaquettes ($n$), influenced by the edge orientation. Specifically, for armchair NRs:

\begin{equation}
\label{equ2}
W(n) = \sqrt{3}d_1\left(n - \dfrac{1}{2}\right),
\end{equation}

\noindent and for zigzag edges:

\begin{equation}
\label{equ3}
W(n) = \dfrac{d_1}{2}(3n - 1).
\end{equation}

\begin{figure}[h!]
    \centering
        \includegraphics[width=1\columnwidth]{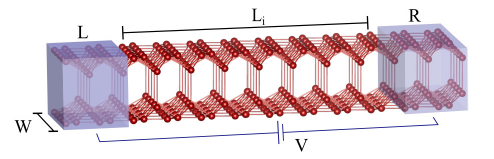}
    \caption{(a) Bilayer borophene transport configuration. The scattering region of length $L_i$ and width $W$ is attached to semi-infinite leads. Zigzag (b) and armchair (c) edges are shown. The width $W$ is measured by the number of primitive cells $n$.}
    \label{fig4}
\end{figure}

The electronic and transport properties were calculated using the \textit{Kwant} package~\cite{groth_kwant_2014}. For instance, the electronic band structure was obtained using the \texttt{kwant.physics.Bands} function. To compute the conductance, defined as \( G(E) = G_0 \operatorname{Tr}\left[t^{\dagger} t\right] \) with \( G_0 = \frac{2e^2}{h} \), the system was divided into three regions (as depicted in Fig.~\ref{fig4}): two semi-infinite $\mathcal{BB}$ nanoribbons attached to the left (L) and right (R) leads, and a central scattering region of length \( L_i \). The transmission from the left lead (0) to the right lead (1) was computed using the \texttt{kwant.solvers.common.Smatrix} and \texttt{Smatrix.transmission(0,1)} functions. Note that, since this study operates within a ballistic transport framework where phase coherence is maintained, the lengths \( L_{\mathrm{AC}} = 1.89\,\text{nm} \) and \( L_{\mathrm{ZZ}} = 1.14\,\text{nm} \) were kept constant throughout the systems.

%%%%%%%%%%%%%%%%%%%%%%%%%%%%%%%%%%%%%%%%%%%%
%%%%%%%%%%%%%%%%%%%%%%%%%%%%%%%%%%%%%%%%%%%%

\section{Results and Discussion\label{sec3}}

%%%%%%%%%%%%%%%%%%%%%%%%%%%%%%%%%%%%%%%%%%%%
\subsection{Periodic Boundary Conditions\label{sec3.1}}
%%%%%%%%%%%%%%%%%%%%%%%%%%%%%%%%%%%%%%%%%%%%

\begin{figure}[h!]
    \centering
    \includegraphics[width=\columnwidth]{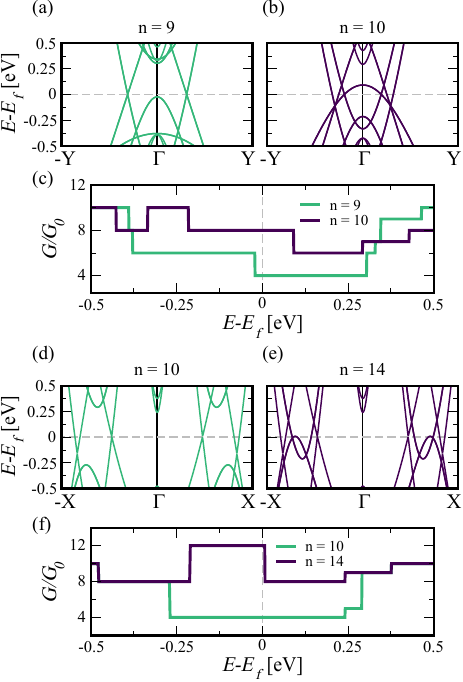}
    \caption{Band structures for armchair PBC nanoribbons with unit cell widths of \( n = 9 \) (a) and \( n = 10 \) (b), and their respective conductance plots (c). Band structures for zigzag PBC nanoribbons with unit cell widths of \( n = 10 \) (d) and \( n = 14 \) (e), and their respective conductance plots (f). $G_{0} = 2e^{2}/h$.}
    \label{fig5}
\end{figure}

To prevent interference from edge states~\cite{nakhaee_dirac_2018} in the initial analysis of the NR properties, periodic boundary conditions were initially applied along the width direction. For narrow armchair NRs ($n=9$) at low energies, as shown in Fig.~\ref{fig5}(a), the band structure features two well-defined cones symmetrically positioned on both sides of the center of the Brillouin zone ($\Gamma$). Interestingly, increasing the width by just one hexagonal plaquette leads to a rapid change in the band structure. For $n=10$, as shown in Fig.~\ref{fig5}(b), four cones emerge, but the cones are shifted, and their intersection is no longer at \( E = 0 \) eV. Additionally, bands with quadratic dispersion are present. This trend is not unique to armchair edges; zigzag NRs in Fig.~\ref{fig5}(d) and (e) also show significant variation. For instance, the $n=10$ case exhibits four cones, while the band structure for $n=14$ features eight cones, accompanied by quadratic bands. The strong dependence of the electronic properties of the NRs on width is also evident in the conductance, as shown in Fig.~\ref{fig5}(c) for armchair NRs and panel (f) for zigzag NRs. While conductance quantization is clearly observed, the conductance at \( E = 0 \) eV does not show the same value across different widths. Additionally, new plateaus appear as the width increases.

%%%%%%%%%%%%%%%%%%%%%%%%%%%%%%%%%%%%%%%%%%%%
%%%%%%%%%%%%%%%%%%%%%%%%%%%%%%%%%%%%%%%%%%%%

\begin{figure}[h!]
    \centering
    \includegraphics[width=\columnwidth]{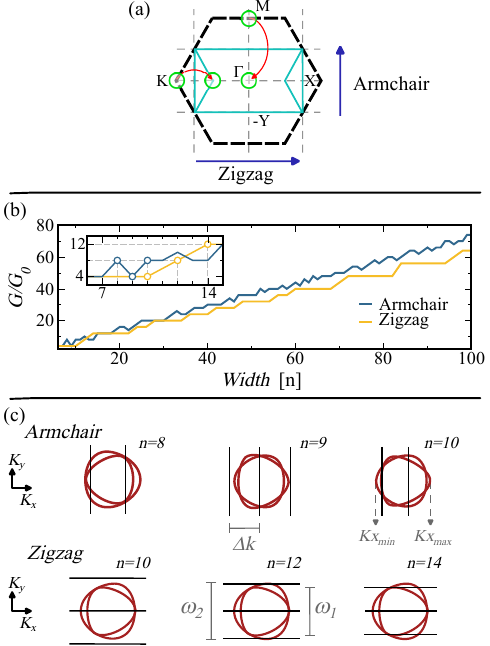}
    \caption{(a) The first hexagonal Brillouin zone (black) of bilayer borophene is folded into its first rectangular Brillouin zone (blue). High-symmetry points are displaced, and new reference points appear, \( X \) and \( Y \), with the new folded paths shown in (c).(b) Conductance for armchair and zigzag PBC nanoribbons at the Fermi energy as the width of the unit cell varies. \( G_{0} = 2e^{2}/h \).  (c) Folding crosses through bulk energy bands at Fermi energy (red) for highlighted sizes  with armchair and zigzag PBC nanoribbons.}
    \label{fig6}
\end{figure}

To understand the impact of the nodal line and its relation to the width's effect on electronic properties, Fig.~\ref{fig6}(a) illustrates the hexagonal Brillouin zone of bulk $\mathcal{BB}$, along with its folding and projection onto the armchair and zigzag directions. First, the $K$-point projects onto the center of the Brillouin zone for armchair NRs, while for zigzag NRs, it projects onto \( k = \pm 2\pi/3 \). This suggests that energy points near \( E \sim 0 \) eV with linear dispersion are projections of the nodal line. 

%%%%%%%%%%%%%%%%%%%%%%%%%%%%%%%%%%%%%%%%%%%%
%%%%%%%%%%%%%%%%%%%%%%%%%%%%%%%%%%%%%%%%%%%%

Second, due to the inverse relationship between real and reciprocal spaces, as the width \( W \) increases in real space, the transverse momentum lines in reciprocal space become closer, covering a larger portion of the Brillouin zone. To examine the effect of cutting and projecting the nodal line, Fig.~\ref{fig6}(b) shows conductance as a function of \( W \) for both types of edges. Zigzag NRs (yellow) exhibit a stepped behavior, with conductance increasing by \( 2G_0 \) and \( 4G_0 \), interrupted by plateaus of constant conductance. Armchair NRs (blue), however, show a more erratic pattern. Conductance increases of up to \( 4G_0 \) are followed by sudden drops of equal magnitude, as seen in the inset. Despite this irregularity, the conductance shows an overall increasing trend with a slope similar to that of zigzag nanoribbons.

To understand the origin of the conductance oscillations, Fig.~\ref{fig6}(c) shows the transverse momentum planes, spaced by \( \Delta k = 2\pi /W \), plotted in black around the \( K \)-point of the 2D $\mathcal{BB}$ structure for three different widths. Additionally, the constant energy contours of the Dirac cones at \( E=0 \) eV are displayed in red. The nodal line arises from the crossing of two Dirac cones, so the contour lines appear as pairs of rings, each corresponding to one cone.  

For armchair NRs with \( n = 8 \), four crossings are identified in each ring, leading to a total of eight crossings from this \( K \)-point. Considering the other non-equivalent \( K' \)-point, there are a total of sixteen crossings. Since only half of these crossings have a positive slope, the expected conductance is \( 8G_0 \), which aligns with the results shown in Fig.~\ref{fig6}(b). For \( n = 9 \), only one momentum plane intersects the nodal line, resulting in the loss of four transport channels at \( E_F = 0 \) eV compared to the previous case. However, for \( n = 10 \), two intersecting planes are present again, restoring the conductance to \( G = 8G_0 \). In contrast, for zigzag NRs, the momentum planes align with the center of the Dirac cones at the \( K \)-point, and the number of crossings is never reduced. This results in constant conductance values or steps of \( 4G_0 \) as a function of \( W \).
%%%%%%%%%%%%%%%%%%%%%%%%%%%%%%%%
%%%%%%%

The geometric approach outlined above allows for the determination of the conductance at \(E_F = 0\) eV, where the number of modes corresponds to the intersections between the nodal line and the transverse momentum planes. This can be approximated by dividing the diameter of the nodal line by \(\Delta k = 2\pi/W(n)\). However, this linear expression in \(n\) is only valid for wide NRs. The correct expression must consider that the nodal line is not flat and exhibits different trigonal warping effects for the valence and conduction bands \cite{nakhaee_dirac_2018}. Taking all these factors into account, the conductance can be correctly estimated for armchair NRs as follows:

\begin{equation}
    G_{\mathrm{armchair}}(n) 
    \,=\,
    \left\lfloor 
        0.4284\,\frac{d_1\,(3n - 1)}{4\pi}
    \right\rfloor
    \;+\;
    \left\lfloor 
        0.3628\,\frac{d_1\,(3n - 1)}{4\pi}
    \right\rfloor,
\end{equation}

\noindent and for zigzag NRs:

\begin{equation}
\begin{split}
G_{\mathrm{zigzag}}(n) 
\,=\,&
4\,\Bigl(
\bigl\lfloor 
      0.2142\,\frac{\sqrt{3}\,d_1\,(n - 0.5)}{2\pi}
 \bigr\rfloor 
\\
&\;+\;
\bigl\lfloor 
      0.1814\,\frac{\sqrt{3}\,d_1\,(n - 0.5)}{2\pi}
 \bigr\rfloor
\Bigr).
\end{split}
\end{equation}

In these expressions, $\bigl\lfloor x \bigr\rfloor$ represents the floor function, which returns the integer part of \( x \). The numerical constants \(0.4284\), \(0.3628\), \(0.2142\), and \(0.1814\) correspond to the effective widths of the two trigonal loops into which the nodal line can be decomposed. These effective parameters differ for the armchair and zigzag directions, as presented in the equations.

%%%%%%%
%%%%%%%%%%% 
\begin{figure}[t]
    \centering
    \includegraphics[width=\columnwidth]{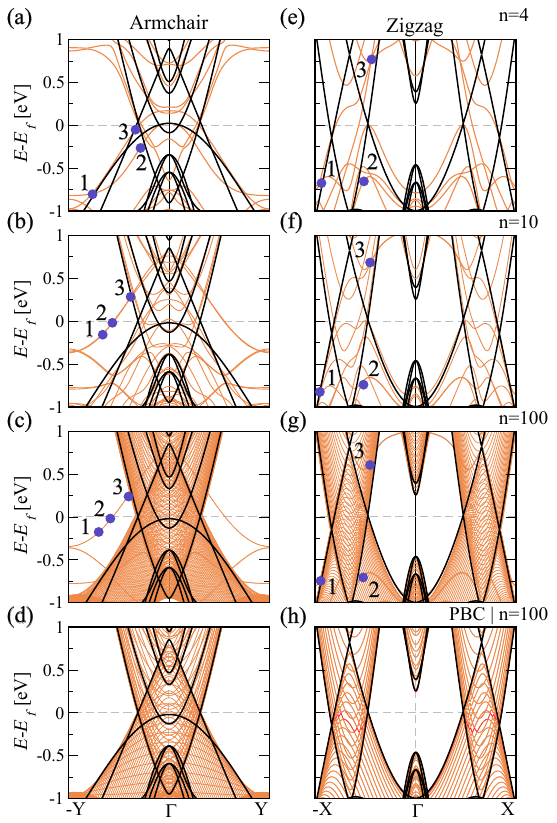}
    \caption{Comparative band structures for armchair [(a)--(c)] and zigzag [(e)--(g)] edged nanoribbons with widths $n = 4$, $n = 10$, and $n = 100$, and armchair (d) and zigzag (h) PBC nanoribbons with width $n = 100$. Black lines represent the idealized bulk band structure, accounting for the effects of the first Brillouin zone folding.}
    \label{fig7}
\end{figure}

\subsection{Hardwall boundary conditions and edge states\label{sec3.2}}

After examining the electronic dispersion and conductance of zigzag and armchair ($\mathcal{BB}$) NRs under periodic boundary conditions, we now turn our attention to the influence of edge states. Fig.~\ref{fig7} presents the band structures of armchair and zigzag NRs with hardwall boundary conditions for widths $n = 4$, $n = 10$, and $n = 100$. For comparison, in all panels the band structure of bulk 2D $\mathcal{BB}$ along the armchair and zigzag directions is plotted in black, alongside the band structure with periodic boundary conditions for $n = 100$ in Figs.~\ref{fig7}(d) and (h).This comparison enables the identification of the edge state band for wider NRs as those bands that fall outside the region defined by the PBC bands: for armchair terminations, between \(-0.5\)~eV and \(0.5\)~eV, and for zigzag NRs, between \(-0.7\)~eV and \(0.7\)~eV. This position is consistent with previous studies, where edge states have been reported to be energetically distant from the Fermi level~\cite{nakhaee_tight-binding_2020, gao_structure_2018}. For narrower NRs ($n \leq 10$), the edge state band shifts to higher energies, and the cones seen under periodic boundary conditions —resulting from the cut and projection of the nodal line— disappear, indicating strong coupling between the edge states.

%%%%%%%%%%%%%%%%%
\begin{figure}[t]
    \centering
    \includegraphics[width=\columnwidth]{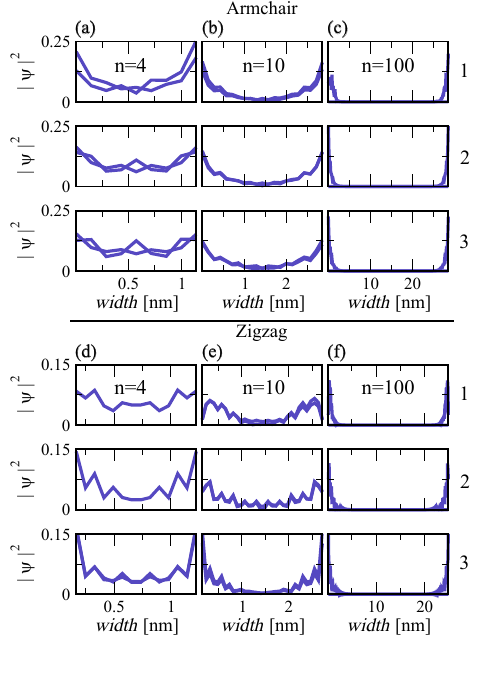}
    \caption{Squared wave function amplitude $|\psi|^2$ for bilayer borophene nanoribbons with armchair edges [(a)--(c)] and zigzag edges [(d)--(f)] for widths $n = 4$, $n = 10$, and $n = 100$. Edge states are localized for narrow ribbons and decouple from the bulk as width increases.}
    \label{fig8}
\end{figure}
%%%%%%%%%%%%%%%%%

To evaluate the degree of localization and coupling between states on opposite edges, Fig.~\ref{fig8} shows the probability density for various $k$-values. The columns correspond to specific ribbon widths, while the first, second, and third rows display $|\Psi|^2$ plots for the $k$-values marked by the blue dots labeled as $1$, $2$, and $3$ in the band structure (Figs.~\ref{fig7}). For a given width, there is no significant variation with respect to the $k$-values, as the penetration depth and height of the edge states remain largely unchanged as we move down the column. On the other hand, as we move to the right across a row, increasing the width of the NRs, we observe that in very narrow NRs ($n = 4$), there is a strong overlap of edge states in the middle of the ribbon. The coupling between edge states decreases rapidly, and by $n = 10$, only a small overlap is observed.  The  density plot presented for the zigzag NR with $n=10$ and labeled by the dot $1$, presents a M-shape pattern. Although, there is a non-zero density at the edges  the highest values very close to it producing a the M-pattern . This case was selected to show the richness of the edge states in narrow $\mathcal{BB}$ NRs. For wider NRs, the coupling becomes nearly nonexistent, with the edge states appearing fully localized at the edges, as shown for $n = 100$. While not shown here, it is important to note that the edge states have nearly equal contributions from all four orbitals, and the plots represent the sum of these contributions at each atomic site. 

%%%%%%%%%%%%%%%%%%%%%%%%
\begin{figure}[t!]
    \centering
    \includegraphics[width=\columnwidth]{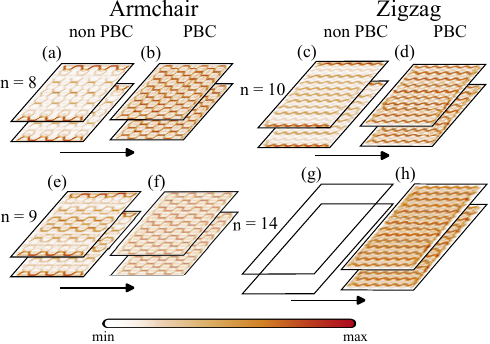}
    \caption{Current density at the Fermi energy (0 eV) for armchair nanoribbons of width \( n = 8 \) (a) and \( n = 9 \) (c), and for zigzag nanoribbons of width \( n = 10 \) (b) and \( n = 14 \) (d). For each size it is considered both with and without PBC at $E=0$ eV.}
    \label{fig9}
\end{figure}
%%%%%%%%%%%%%%%%%%%%%%%%

\begin{figure}[t]
    \centering
    \includegraphics[width=\columnwidth]{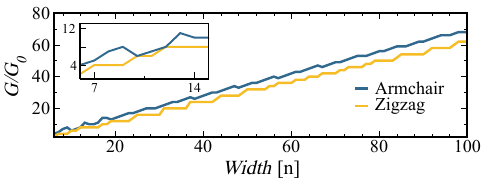}
    \caption{Conductance at Fermi energy as a function of nanoribbon width, measured in terms of the number of cells. The blue line corresponds to the armchair nanoribbon, and the yellow line corresponds to the zigzag nanoribbon.}
    \label{fig10}
\end{figure}

Unlike the edge states seen in monolayer zigzag nanoribbons (NRs)~\cite{gupta_dirac_2018}, the edge state bands for zigzag and armchair $\mathcal{BB}$ NRs are not flat, and a current density can be carried by those modes. In narrow NRs, the band is not detached from other bands~\cite{paez_zigzag_2016} and high current density is observed at the edges and non-zero values in the middle, as shown in Figs.~\ref{fig9}(a) and (c). In wider NRs, edge states have a reduced impact on transport, as the current pattern is inverted, with lower current at the edges and higher density in the center, primarily carried by bulk states. This is evident when comparing the current density of NRs with hardwall boundary conditions to their counterparts with periodic boundary conditions, as shown in Fig.~\ref{fig9}. Further evidence is provided in Fig.~\ref{fig10}, which plots the conductance at $E=0$ eV as a function of ribbon width. The blue line represents the armchair, while the yellow line corresponds to the zigzag configuration. Although the overall trend is qualitatively similar to the case with periodic boundary conditions, specific values for conductance oscillations and plateau heights differ. These values must be directly calculated from the conductance, as the presence of edge states prevents using the cut-and-projection method to determine them.

%%%%%%%%%%%%%%%%%%%%%%%%
%%%%%%%%%%%%%%%%%%%%%%%%

%%%%%%%%%%%%%%%%%%%%%%%%
\section{Effect of Strain on electronic and transport properties}
\label{sec_strain}
%%%%%%%%%%%%%%%%%%%%%%%%

The approach we have adopted to gain physical insight into the electronic transport properties of zigzag and armchair $\mathcal{BB}$ nanoribbons (NRs) is based on the cut-and-projection method of the nodal line. Building on this foundation, the next step involves deforming the perfect hexagonal Brillouin zone by applying uniaxial tensile strain. Our objective is to investigate the impact of mechanical deformation on the previously discussed electronic transport characteristics, rather than performing a comprehensive analysis of the effects of strain on the electronic properties of $\mathcal{BB}$ \cite{Shukla_2018,Liu_2022}. To this end, we restrict our study to a single strain value of 5\%, applied along the X and Y direction regardless of the edge type.  Here, the strain is defined as $\epsilon = (a - a_0) / a_0$, where $a$ and $a_0$ represent the lattice constants along X and Y direction  with and without strain, respectively. 

From an operational standpoint, the numerical methodology remains unchanged: the electronic bands of the strained $\mathcal{BB}$ are computed using \textsc{Quantum ESPRESSO}, followed by fitting the tight-binding parameters with \textsc{TBStudio}, which are then used to calculate the transport properties in \textsc{Kwant}. Table \ref{table:TBpara} summarizes the updated parameters for the strained cases.

\begin{figure}[t]
    \centering
    \includegraphics[width=\columnwidth]{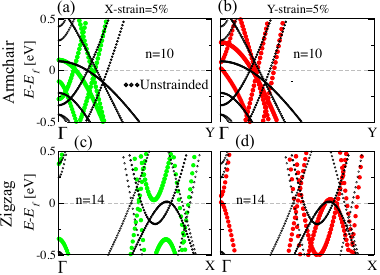}
    \caption{Band structures of bilayer borophene nanoribbons with armchair (top row) and zigzag (bottom row) edge terminations under uniaxial tensile strain. The left column (a, c) shows band structures with 5\% strain along the X-direction, while the right column (b, d) corresponds to 5\% strain along the Y-direction. Strained systems are shown in color (green for X-strain, red for Y-strain), while the unstrained band structures are depicted in black for comparison. 
}
    \label{fig11}
\end{figure}

Figure~\ref{fig11} presents the band structure of strained armchair ($n=10$) and zigzag ($n=14$) NRs with PBC under a tensile strain of 5\%. The results for strain applied along the X-direction (green) are shown in panels (a) and (c), while those for strain along the Y-direction (red) are displayed in panels (b) and (d). Notably, the strain is parallel to the current flow in panels (b) and (c). 

As expected, strain significantly modifies the band structure compared to the unstrained case, represented by black dots. However, unlike Dirac point semimetals, no gap is observed \cite{Pereira_2009}. The projection of the nodal line is evident from the presence of linear dispersing bands, but the position in momentum space and the arrangement of the bands differ from the pristine case. This variation arises from the intersection regularly spaced transverse momenta, which depend on the NR width,  with the shifted cones \cite{Bahamon_2013}.
 
 %%%%%%%%%%%%%%%%%%%%%%%%
%%%%%%%%%%%%%%%%%%%%%%%%

\begin{figure}[t!]
    \centering
    \includegraphics[width=\columnwidth]{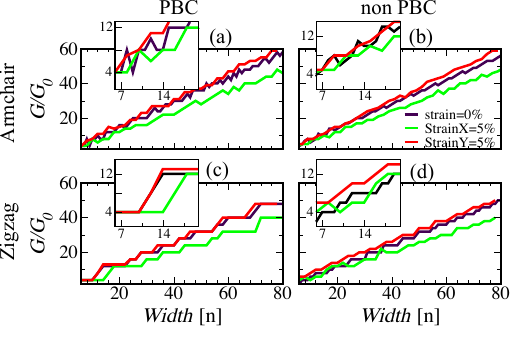}
    \caption{Conductance ($G/G_0$) as a function of the width ($W$) for bilayer borophene nanoribbons with armchair and zigzag edge terminations under periodic boundary conditions (a,c panel) and non-periodic boundary conditions (b,d panels). The conductance was evaluated for unstrained (black), 5\% strain along the $x$-direction (green), and 5\% strain along the $y$-direction (red) configurations. A stepwise conductance pattern is observed, with plateaus interrupted by increasing conductance steps. Armchair nanoribbons display a more irregular conductance trend compared to the smoother progression seen in zigzag configurations. Insets highlight the finer conductance steps in narrower ribbons.
}
    \label{fig12}
\end{figure}

Having demonstrated the effect of strain on the band structure, we now present in Fig.~\ref{fig12} the conductance at the charge neutrality point for strained NRs with and without PBC as a function of width. It is evident that strain modifies the number of available transport channels. When strain is applied along the X-direction, the number of channels decreases compared to the pristine case, whereas it increases for strain applied along the Y-direction.  As in the unstrained case, narrow armchair NRs exhibit oscillations, as shown in the insets of panels (a) and (b). For zigzag NRs, the conductance behaves more uniformly, increasing in steps of $4G_0$, similar to the unstrained case. However, strain alters the initial conductance values, reflecting its influence on the electronic transport properties of nodal line semimetals.

%%%%%%%%%%%%%%%%%%%%%%%%
\section{Final Remarks\label{sec4}}
%%%%%%%%%%%%%%%%%%%%%%%%

It is widely recognized that the electronic properties of materials are influenced by the shape of the Fermi surface; however, it is essential to understand the specific distinctions involved. In this work, the case of $\mathcal{BB}$ was examined, where the conduction and valence bands meet along a loop in momentum space. This configuration results in a distinct conductance behavior in nanoribbons compared to that of point-like Dirac materials. These differences are particularly evident around \( E \sim 0 \)~eV, where the conductance increases with width, exhibiting oscillations in narrow NRs even in the presence of edge states. This perspective is further supported by the introduction of a 5\% uniaxial tensile strain, which enables the engineering of conductance by either increasing or decreasing the number of available channels.   It is worth noting that a tight-binding model was developed to explore these properties and can be utilized in future studies to investigate the effects of disorder or magnetic fields with or without strain.

%%%%%%%%%%%%%%%%%%%%%%%%
%%%%%%%%%%%%%%%%%%%%%%%%

\section*{Data availability}
Data will be made available on request.

% ==================================================
\section*{Acknowledgements}

C.J.P.G. acknowledges the support of the Departamento Administrativo de Ciencia, Tecnología Innovación, Colombia, and Ministerio de Ciencia Tecnología e Innovación, Colombia, through the announcement No 852-2019 and financing contract No. 80740-535-2020 (Project ID: 1102-852-69674) and the Universidad Industrial de Santander (UIS), Colombia, (Project ID: 9483-2666).  D. A. B. acknowledges support from CAPES-PRINT (88887.310281$/$2018-00),  CNPq (309835$/$2021-6) and Mackpesquisa. DAB is also grateful to the hospitality of School of Physics-UIS, where this work was initiated.

%===========================================================

\end{document}